\documentclass[fleqn,usenatbib]{mnras}

\usepackage{newtxtext,newtxmath}

\usepackage[T1]{fontenc}
\usepackage{ae,aecompl}

\usepackage{graphicx}	
\usepackage{amsmath}	
\usepackage{amssymb}	
\usepackage{lscape}
\usepackage{rotating}
\usepackage{color}

\title{Polarization studies of Rotating Radio Transients}

\author[M.~Caleb et al.]
{M.~Caleb$^{1, 2, 3}$\thanks{Email: manisha.caleb@manchester.ac.uk},
W.~van Straten$^{4,3}$,
E. F.~Keane$^{5,3}$, 
A.~Jameson$^{6,3}$,
M.~Bailes$^{6,3}$,
\newauthor E. D.~Barr$^{7,3}$,
C.~Flynn$^{3}$,
C. D.~Ilie$^{1}$,
E.~Petroff$^{8,3}$,
A.~Rogers$^{4}$,
B. W.~Stappers$^{1}$,
\newauthor V.~Venkatraman Krishnan$^{7,3}$,
P.~Weltevrede$^{1}$
\\ \\
$^{1}$ University of Manchester, Jodrell Bank Centre of Astrophysics, Alan Turing Building, Manchester M13 9PL, UK \\
$^{2}$ Research School of Astronomy and Astrophysics, Australian National University, ACT, 2611, Australia\\
$^{3}$ Centre for Astrophysics and Supercomputing, Swinburne University of Technology, P.O. Box 218, Hawthorn, VIC 3122, Australia \\
$^{4}$ Institute for Radio Astronomy \& Space Research, Auckland University of Technology, Private Bag 92006, Auckland 1142, New Zealand \\
$^{5}$ SKA Organisation, Jodrell Bank Observatory, SK11 9DL, UK \\ 
$^{6}$ OzGrav, Centre for Astrophysics and Supercomputing, Swinburne University of Technology, H11 PO Box 218 Hawthorn, Vic, 3122, Australia \\ 
$^{7}$ Max-Planck-Institut f\"ur Radioastronomie, Auf dem H\"ugel 69, D-53121 Bonn, Germany \\
$^{8}$ Anton Pannekoek Institute for Astronomy, University of Amsterdam, Science Park 904, 1098 XH Amsterdam, The Netherlands}

\date{Accepted XXX. Received YYY; in original form ZZZ}

\pubyear{2019}

\begin{document}
\label{firstpage}
\pagerange{\pageref{firstpage}--\pageref{lastpage}}
\maketitle

\begin{abstract}

We study the polarization properties of 22 known rotating radio transients (RRATs) with the 64-m Parkes radio telescope and present the Faraday rotation measures (RMs) for the 17 with linearly polarized flux exceeding the off-pulse noise by 3$\sigma$. Each RM was estimated using a brute-force search over trial RMs that spanned the maximum measurable range $\pm1.18 \times 10^5 \, \mathrm{rad \, m^2}$ (in steps of 1 $\mathrm{rad \, m^2}$), followed by an iterative refinement algorithm. The measured RRAT RMs are in the range |RM| $\sim 1$ to $\sim 950$ rad m$^{-2}$ with an average linear polarization fraction of $\sim 40$ per cent. Individual single pulses are observed to be up to 100 per cent linearly polarized. The RMs of the RRATs and the corresponding inferred average magnetic fields (parallel to the line-of-sight and weighted by the free electron density) are observed to be consistent with the Galactic plane pulsar population. Faraday rotation analyses are typically performed on accumulated pulsar data, for which hundreds to thousands of pulses have been integrated, rather than on individual pulses. Therefore, we verified the iterative refinement algorithm by performing Monte Carlo simulations of artificial single pulses over a wide range of S/N and RM. At and above a S/N of 17 in linearly polarized flux, the iterative refinement recovers the simulated RM value 100 per cent of the time with a typical mean uncertainty of $\sim5$ rad m$^{-2}$. The method described and validated here has also been successfully used to determine reliable RMs of several fast radio bursts (FRBs) discovered at Parkes.

\end{abstract}

\begin{keywords}
magnetic fields - polarization - methods: data analysis - surveys - radio continuum: transients
\end{keywords}



\section{Introduction}

Over the last two decades, advances in radio astronomy instrumentation and signal processing software have enabled high time resolution studies of the dynamic radio sky and opened up the largely unexplored domain of fast radio transients \cite[e.g.][]{Cordes2003,PALFA,Keith}.
These developments led to the discovery of two similar but fundamentally different classes of radio transients: the Galactic Rotating Radio Transients (RRATs) and the extragalactic Fast Radio Bursts (FRBs).
The RRATs were first discovered in 2006, in archival data from the
Parkes Multibeam Pulsar Survey \citep{McLaughlin}.
Whereas the large majority of pulsars have been discovered through Fourier domain searches for periodicity, RRATs are detected as single pulses with widths between 0.3 -- 100 ms and peak flux densities in the range 0.01 -- 3.6 Jy\footnote{\url{http://astro.phys.wvu.edu/rratalog/}}.
Since no periodic emission from RRATs is detected, it has been suggested that RRATs are either weakly emitting or not emitting (i.e. nulling) during the periods between detected pulses. If such non-detections are due to them being off/nulling, then their population is possibly greater than or equal to that of normal radio pulsars \citep{KeaneKramer}. The difficulty in detecting them due to their variability, implies a largely undetected source population \citep{KeaneKramer} which does not comply with the estimated core-collapse supernovae rate in the Milky Way, thereby causing a discrepancy.
This over-estimation of the RRAT population could be due to extrapolations from a small number of known sources\footnote{\url{http://astro.phys.wvu.edu/rratalog/}} and can be reconciled if they somehow form a particular stage in pulsar evolution \citep{Keane_rrats}. 

The most common definition of an RRAT is a neutron star that is much more readily detectable in single pulse searches than in periodicity searches. Consequently, the term RRAT is more closely related to a source's detectability than to the actual physical mechanism of emission \citep{Keane2010, Weltevrede}. Their underlying periodicities are typical of slow pulsars, lying between 0.1 -- 7.7 seconds, and their pulse amplitude distributions are log-normal \citep{keane_pulseamp, Cui} with time intervals between single pulses lasting anywhere
between seconds ($\sim7$ pulses per minute) to hours ($\sim2$ pulses per hour) \citep{Keane_rrats}. RRATs typically have peak luminosities of $\sim10$ Jy kpc$^{2}$ and high brightness temperatures of $10^{22} - 10^{23}$ K, which indicate coherent non-thermal emission with a causally connected emission region. 

Magnetic fields play pivotal roles in many aspects of astrophysics, ranging from star formation to galactic dynamics \citep{Boulanger}. They make a significant contribution to the hydrostatic balance in the interstellar medium \citep{Widrow}, yet much remains unknown about how these fields are generated or how they are evolving. Polarization studies are what is required to answer 
these questions of magnetogenesis. For example, the magnetic field of the Galaxy has been studied through (polarized) synchrotron emission \citep{Dickey}, optical starlight polarization \citep{Hall, SAMao}, Zeeman splitting of maser lines \citep{Momjian} and Faraday rotation \citep{Han,Noutsos}. 
Faraday rotation of polarized radio sources 
has proven to be a powerful probe of the Galactic magnetic field by providing a measure of the strength and direction of the line-of-sight component of the field. 
Pulsars are well distributed throughout the Galactic disk and typically have substantial linear 
polarization, thereby making Faraday rotation relatively easy to measure. Over the last couple of decades, the Galactic magnetic field has been mapped through polarization studies of pulsars, and
these remarkable maps have provided the most reliable evidence for a 
clockwise-directed field (as viewed
from the North Galactic pole) in the inter-arm regions and a counterclockwise-directed field along the arm regions \citep{Han,RBeck,Han2018, Sobey}.

The first measurements of radio polarization and Faraday rotation of an RRAT were made by \cite{Karastergiou} for J1819$-$1458.
These measurements were made using 72 single pulses, which were observed over multiple epochs and integrated into a 
single stable profile. 
In this paper, we study the polarization properties of 22 RRATs via sporadic single pulses and compare these with that of an integrated profile, and with the polarization properties of the existing pulsar population.
We also discuss the reliability of using this to study the polarization properties of FRBs which are a new class of millisecond duration, coherent radio emission of extragalactic origin \citep{Petroff_review}. 
The layout of the paper is as follows: In Section \ref{sec:FR} we explain the phenomenon of Faraday rotation followed by our observations and analyses in Section \ref{sec:obs}. We present our results in Section \ref{sec:RandD} and finally our discussion and conclusions in Section \ref{sec:conc}.

\section{Faraday rotation}
\label{sec:FR}

Linearly polarized light can be represented as the superposition of two circularly polarized components of opposite hand. On propagation through a magnetised cold plasma, the phase velocities of 
the two circularly polarized electric field components differ, equivalently rotating the plane of polarization of linearly polarized light; this observable effect of birefringence is known as Faraday rotation.

The polarization of electromagnetic radiation is conventionally described using the Stokes parameters, which include I (total intensity), Q and U (linear polarization) and V (circular polarization). The total amount of linearly polarized flux is given by $L = \sqrt{Q^2 + U^2}$ and the position angle (PA) of the linearly polarized component,
\begin{equation}
    \psi = 0.5 \, \mathrm{tan^{-1}}\Bigg(\frac{U}{Q}\Bigg).
\end{equation}

\cite{Gould} studied the integrated pulse profiles of a sample of 300 pulsars at 600 and 1400 MHz and determined the average degree of linear polarization $\langle L/I \rangle$ to be $\sim20$ per cent and the average degree of circular polarization $\langle |V|/I \rangle$ to be $\sim 10$ per cent at the two frequencies. Considerable variation from pulsar to pulsar is observed around these mean values with individual pulses being up to 100 per cent linearly polarized. If the emission from RRATs has a similar degree of polarization, then it will be straightforward to measure the Faraday rotation occurring in the ISM along the lines-of-sight to them.

Owing to Faraday rotation, the observed PA is given by
\begin{equation}
\psi = \psi_{0} + \mathrm{RM}\lambda^{2},
\end{equation}
where $\psi_{0}$ is the intrinsic PA (assumed to be frequency-independent) of the source and $\lambda$ is the wavelength in metres. The rotation measure (RM; in radians per square meter) quantifies the amount of Faraday rotation and is given by
\begin{equation}
\label{eq:RM}
    \mathrm{RM} = \frac{e^{3}}{2\pi m_\mathrm{e}^{2} c^{4}} \int_0^D \! n_{e} \, \boldsymbol{B} \cdot \boldsymbol{dl},
\end{equation}
%
where $n_e$ is the electron density in particles per cubic centimetre, $\boldsymbol{B}$ is the magnetic field in microgauss and $\boldsymbol{dl}$ is the elemental vector towards the observer along the line-of-sight in parsecs.
When combined with the dispersion measure (DM) given by
\begin{equation}
\label{eq:DM}
\mathrm{DM} = \int_0^D \! n_{e} \, dl,
\end{equation}
the RM can be used to estimate the average component of the magnetic field parallel to the line-of-sight weighted by the local free electron density \citep{Lyne},
\begin{equation}
\label{eq:Bll}
\langle B_{\parallel} \rangle \simeq 1.232 \, \bigg(\frac{\mathrm{RM}} {\mathrm{rad \, m^{-2}}} \bigg) \bigg(\frac{\mathrm{DM}} {\mathrm{pc \, cm^{-3}}} \bigg)^{-1} \; \upmu\mathrm{G}.
\end{equation}

\noindent The above equation assumes that the magnetic field is constant along the line-of-sight with no field reversals.
Magnetic fields have been mapped in nearby galaxy disks, out to a few $\times$ 10 Mpc \citep{Widrow} and constraints on the IGM magnetic field ($\sim$ nG) can be made from high energy charged cosmic rays \citep{Aharonian, Batista}. Also, nG scale magnetic fields in the IGM (upper limit) are constrained by Faraday rotation measurements of high-$z$ radio loud QSOs \citep{PKronberg}. 
More recently, microgauss strength magnetic fields have been estimated in compact astrophysical sources out to redshifts of $z \sim 1$ \citep[e.g.][]{Mao}. 
However, it should be noted that intervening dense clumps and isolated H\textsc{ii} regions along the lines-of-sight could contribute significantly to the overall observed RM. 


\begin{table*}
\caption{Observed and inferred parameters for the 22 RRATs. Columns: (1) RRAT name based on J2000 coordinates; (2,3) Galactic longitude ($l$) and latitude ($b$) from the RRATalog; (4) dispersion measure \citep{KeaneDMs, SarahDMs, DenevaAO}; (5) inverse-variance weighted average of the single pulse RM distribution $\pm$ its 1-$\sigma$ error; (6) the number of pulses in the distribution; (7) RM of the integrated pulse obtained by adding all the single pulses in the distribution; (8) integrated magnetic field along the line-of-sight calculated from Equation \ref{eq:Bll} using the weighted average RM in column 5; and (9) average linear polarization fraction of the single pulses. The sources marked with $^{\dagger}$ exhibit orthogonally polarized modes in their single pulses and those marked with * show diverse single pulses with varying PAs, thereby preventing us from forming an integrated pulse as the PA points would add incoherently.} 
\centering
\begin{tabular}{l c c c c c c c c}
\hline\hline

J2000 name  & $l$ & $b$ & DM & \multicolumn{2}{c}{Single pulse distribution}  & Integrated pulse & $\langle B_{\parallel} \rangle$ & $\langle L/I \rangle$\\ [0.5ex] 
            & (deg) & (deg) & ($ \mathrm {pc} \, \mathrm{cm^{-3}}$) & RM ($\mathrm{rad} \, \mathrm m^{-2}$) & Pulses & RM ($\mathrm{rad} \, \mathrm m^{-2}$) & $\upmu$G & \% \\
\hline
J0410--31$^{\dagger}$ & 230.59 & $-46.67$ & 9.2 & $16\pm3$ & 9 & $13 \pm 1$ & $2.1\pm0.2$ & $\sim60$\\
J0628+09 & 202.19 & $-0.86$ & 88.0 & $124\pm9$ & 6 & $121\pm 5$ & $1.7\pm0.1$ & $\sim 50$\\
J0837--24 & 247.45 & 9.80 & 142.8 & $352\pm8$ & 2 & $360\pm5$ & $3.0\pm0.05$ & $\sim 50$\\
J0941--39 & 267.70 & 9.98 & 78.2 & $-78\pm21$ & 39 & $-87\pm 6$ & $-1.2\pm0.3$ & $\sim 40$ \\ 
J1129--53 & 290.80 & 7.41 & 77.0 & $-19\pm9$ &  25 & -- & $-0.3\pm0.6$ & $\sim 30$\\ 
J1226--3223 & 296.90 & 30.19 & 36.7 & $-63\pm6$ & 8 & $-63\pm2$ & $-2.1\pm0.3$ & $\sim 30$ \\ 
J1317--5759 & 306.43 & 4.70 & 145.4 & $174\pm12$ & 9 & $185\pm11$ & $1.5\pm0.1$ & $\sim 35$ \\
J1423--56*  & 315.29 & 3.87 & 32.9 & $-1\pm14$ & 3 & -- & $-0.04\pm17$  & $\sim 30$  \\ 
J1513--5946 & 319.97 & $-1.70$ & 171.7 & $104\pm11$ & 7 & $110\pm3$ & $0.7\pm0.1$ & $\sim 35$ \\
J1554--5209 & 329.01 & 1.19 & 130.8 & $-139\pm10$ & 31 & $-144\pm7$ & $-1.3\pm0.08$ & $\sim 70$ \\ 
J1707--4417 & 343.04 & $-2.28$ & 380.0 & $-655\pm21$ & 3 & $-656\pm7$ & $-2.1\pm0.05$ & $\sim 50$ \\ 
J1753--38 & 352.25 & $-6.38$ & 168.4 & $123\pm27$ & 7 & $123\pm21$ & $0.9\pm0.3$ & $\sim 30$ \\ 
J1819--1458$^{\dagger}$ & 16.02 & 0.08 & 196.0 & $323\pm7$ & 41 & $321 \pm 5$ & $2.0\pm0.03$ & $\sim 55$ \\
J1826--1419* & 17.4 & $-1.14$ & 160.0 & $122\pm17$ & 3 & -- & $0.9\pm0.2$ & $\sim 50$ \\ 
J1840--1419$^{\dagger}$* & 18.94 & $-4.12$ & 19.4 & $60\pm21$ & 23 & -- & $3.8\pm0.4$ & $\sim 35$ \\
J1854--1557* & 19.02 & $-7.95$ & 160.0 & $-66\pm18$ & 2 & -- & $-0.5\pm0.4$ & $\sim 50$ \\
J1913+1330* & 47.42 & 1.38 & 175.64 & $945\pm11$ & 3 & --  & $6.6\pm0.01$ & $\sim 40$ \\ 
\hline
J0735--62 & 274.74 & -19.21 & 19.4 & -- & 24 & -- & -- & $\sim 20$  \\
J0847--4316 & 263.44 & 0.16 & 292.5 & -- & 15 & -- & -- &  $\sim 20$ \\
J1048--5838 & 287.47 & 0.48 & 69.3  & -- & 1 & -- & -- & $\sim 15$  \\
J1854+0306 & 35.88 & 0.83 & 192.4 & -- & 16 & -- & -- & $\sim 30$ \\
J1909+06 & 40.91 & $-0.93$ & 35.0 & -- & 3 & -- & -- & $\sim 25$ \\ [1ex] 

\hline  
\end{tabular} 
\label{table:params1} 
\end{table*}

\section{Observations and Analyses}
\label{sec:obs}

Observations of 22 RRATs were undertaken as part of the P864\footnote{Data are available at \url{https://data.csiro.au/dap/}} project using the 21-cm multibeam receiver \citep{Staveley-smith} of the 64-m Parkes radio telescope. The sample of sources chosen from the RRATalog\footnote{\url{http://astro.phys.wvu.edu/rratalog/}} has flux densities $S_{1400} \geq 100$ mJy and is visible in the Parkes sky \citep[i.e. $\delta \leq +20$;][]{Shimmins}.  All detections were made in real-time using the 
\textsc{heimdall}\footnote{\url{http://sourceforge.net/projects/heimdall-astro/}} software package and the data
were recorded using the Berkeley Parkes Swinburne Recorder (BPSR) and the fourth generation Parkes digital filterbank
system (PDFB4) backends to enable pulse profile calibration and studies of the single pulse emission properties. The 21-cm Parkes
multibeam receiver has orthogonal linearly polarized receptors with a calibration probe placed at a $45^{\circ}$ angle to the receptors to enable the
injection of a linearly polarized broad-band and pulsed calibration signal.  A calibration of the frequency dependent differential gain and phase of the receiver system was made by recording the pulsed calibration signal on cold sky for 2 minutes prior to each RRAT observation (see Section \ref{sec:DP} for more details). 

BPSR divides the 400~MHz of bandwidth (from 1182 to 1582 MHz) into
1024 channels using a polyphase filterbank, the voltages are square law detected and integrated over both polarizations and time to yield the total intensity sampled with 64 $\upmu$s time resolution and 8 bits per sample.
During the real-time search for RRAT single
pulses, blocks of data are read in overlapping sections of 16.77 seconds and are searched
over trial DM and pulse width, followed by a clustering algorithm that groups multiple detections of candidate pulses together. A single pulse can be detected over a range of DMs close to the true value and these are combined into one before reporting the DM that maximises the signal-to-noise ratio (S/N). The processing of a gulp is ceased if the number of ungrouped candidates exceeds $10^6$, symptomatic of being saturated by radio frequency interference, and the processing of the next gulp begins. Additionally, the data from each of the 13
beams are merged, and detections coincident in time are rejected as radio frequency interference. The following cuts are applied
to further reduce the number of candidates,

\begin{equation} 
\begin{split} 
\label{eq:criteria}
\mathrm{S/N} \geq 10, \\
N_\mathrm{beams} \leq 4, \\ 
W \leq 32.768 \, \mathrm{ms}, \\ 
N_\mathrm{events} (t_\mathrm{obs} - 2\mathrm{s} \rightarrow t_\mathrm{obs} + 2 \mathrm{s}) \leq 5, \\ 
\end{split}
\end{equation} 

\noindent where the $N_\mathrm{beams}$ condition specifies the number of beams a pulse from a typical astrophysical source is expected to be present in, the width criterion ensures only narrow pulses are selected and the final condition mandates there are no more than
five other candidates within a 4 second window around the candidate of interest. A similar set of selection criteria are used for FRB searches at the Parkes radio telescope \citep{Superb1}. 
This processing is done on average in under 10 seconds. 
A ring buffer retains 120 seconds of 8-bit full-Stokes data from all 13 beams while the real-time processing is in progress. If the real-time
\textsc{heimdall} software identifies a candidate matching all the criteria listed in Equation \ref{eq:criteria}, it saves the 8-bit data
in the buffer within a time window,

\begin{equation}
t_{0} - \Delta t \leq t \leq t_{0} + 2\Delta t,
\end{equation}

\noindent where $t_{0}$ is the time at which the event occurred at the highest frequency of the observing band, $t$ is the time elapsed since the start of 
the observation and $\Delta t$ is the dispersive delay across the whole band. The BPSR backend and the real-time single pulse search pipeline were successfully tested using this study and have since been used to discover most of the FRBs at the Parkes radio telescope after 2013 \citep{Petroff_FRB, nat_keane, RaviSci, Bhandari}.

\begin{figure*}
  \sbox0{\begin{tabular}{@{}cc@{}}
    \includegraphics[scale=0.8]{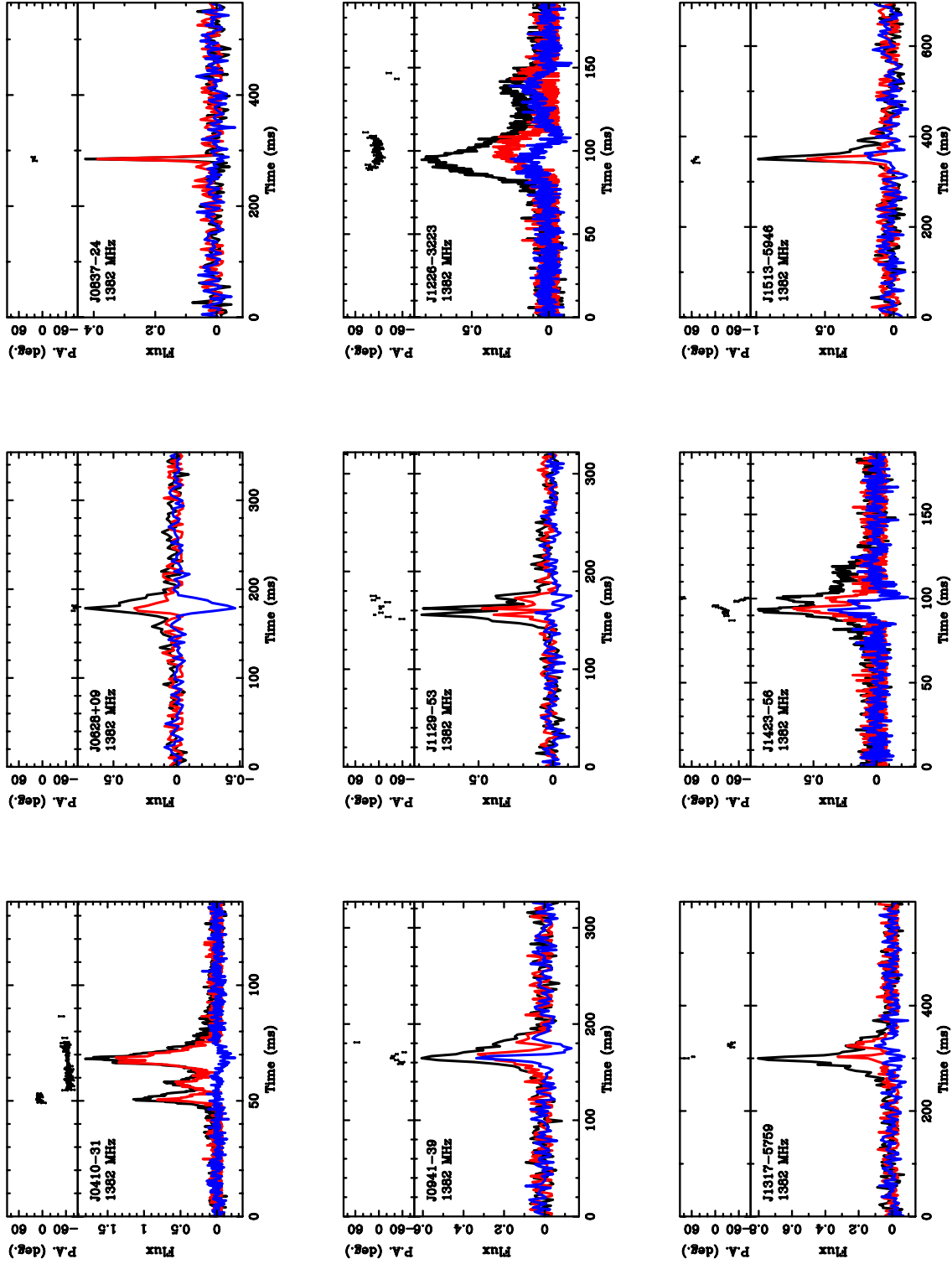}&%
  \end{tabular}}
  \rotatebox{90}{\begin{minipage}[c][\textwidth][c]{\wd0}
    \usebox0
    \caption{Examples of typical polarized single pulse profiles of the RRATs at 1.4 GHz after RM correction. The black, red and blue lines represent the total intensity, linear polarization and circular polarization respectively. The top panel in each plot shows the PA variation as a function of time for those phase bins whose S/N of the linear polarization is $> 3$. The data are not flux calibrated and the flux densities are in arbitrary units.}
  \label{fig:profiles1}
  \end{minipage}}
\end{figure*}

\begin{figure*}
  \sbox0{\begin{tabular}{@{}cc@{}}
    \includegraphics[scale=0.8]{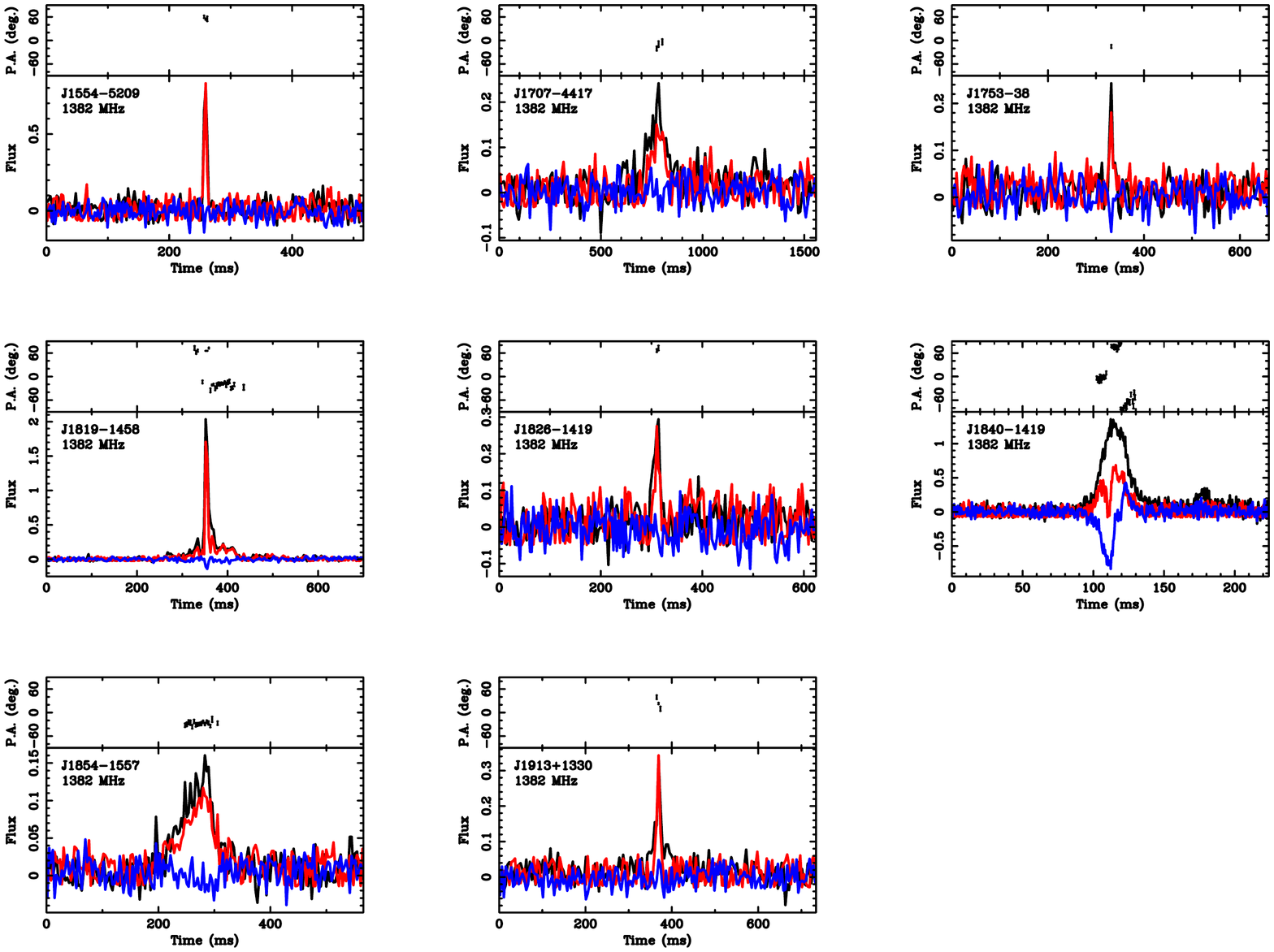}&%
  \end{tabular}}
  \rotatebox{90}{\begin{minipage}[c][\textwidth][c]{\wd0}
    \usebox0
    \caption{Examples of typical polarized single pulse profiles of the RRATs at 1.4 GHz after RM correction. The black, red and blue lines represent the total intensity, linear polarization and circular polarization respectively. The top panel in each plot shows the PA variation as a function of time for those phase bins whose S/N of the linear polarization is $> 3$. The data are not flux calibrated and the flux densities are in arbitrary units.}
\label{fig:profiles2}
  \end{minipage}}
\end{figure*}

\subsection{Offline data processing}
\label{sec:DP}

The offline data processing was performed using the PSRCHIVE \citep{Hotan} pulsar data analysis package. To begin with, the data near the edges ($\sim$15 per cent) of the bandwidth were excised as they are known to be corrupted 
by digitization artefacts. The data are then median smoothed and further cleaned using interactive excision to remove narrow-band radio frequency interference. In keeping with the IAU/IEEE 
convention, for the Parkes 21-cm multibeam receiver we set the symmetry angle to $-{\pi}/2$ \citep{Willem_IAU}. The data are calibrated using the ideal feed assumption\footnote{\url{http://psrchive.sourceforge.net/manuals/pac/}} according to which the receptors are perfectly orthogonally
polarized, and the reference source is 100 per cent linearly polarized and illuminates both receptors equally and in phase. 
The \textsc{rmfit} program was used to perform a brute force search for a peak in the linearly polarized flux $L = \sqrt{Q^{2} + U^{2}}$ as a function of trial RM in the range $\pm1.18 \times 10^5 \, \mathrm{rad \, m^2}$ in steps of 1 $\mathrm{rad \, m^2}$.
This range is defined by the maximum |$\mathrm{RM}$| beyond which the signal would be completely depolarized within a single frequency channel. For each trial RM, \textsc{rmfit} corrects for the Faraday rotation and computes the total linear polarization. A Gaussian is fit to the peak of the ``RM spectrum" and the centroid of this Gaussian is used to determine the best estimate of RM. This first estimate of RM is refined using the algorithm developed by \cite{Han} and described in more detail in Appendix \ref{sec:A1}. First, the data are corrected for Faraday rotation using the RM spectrum centroid value, the band is divided in two and each half is integrated over frequency. The weighted differential position angle between the two halves of the band, $\Delta\Psi$ is then computed. Only those pulse phase bins with linearly polarized flux exceeding the off-pulse noise by 3$\sigma$ in both halves of the band are included in the estimate of differential position angle. If the $\Delta\Psi$ estimate is larger than its uncertainty, then the data are corrected for Faraday rotation with an updated RM estimate and $\Delta\Psi$ is estimated again. This process is repeated until $\Delta\Psi$ is smaller than its uncertainty, at which point the final refined RM is recorded.

\begin{figure}
\centering
\includegraphics[width=3.4 in]{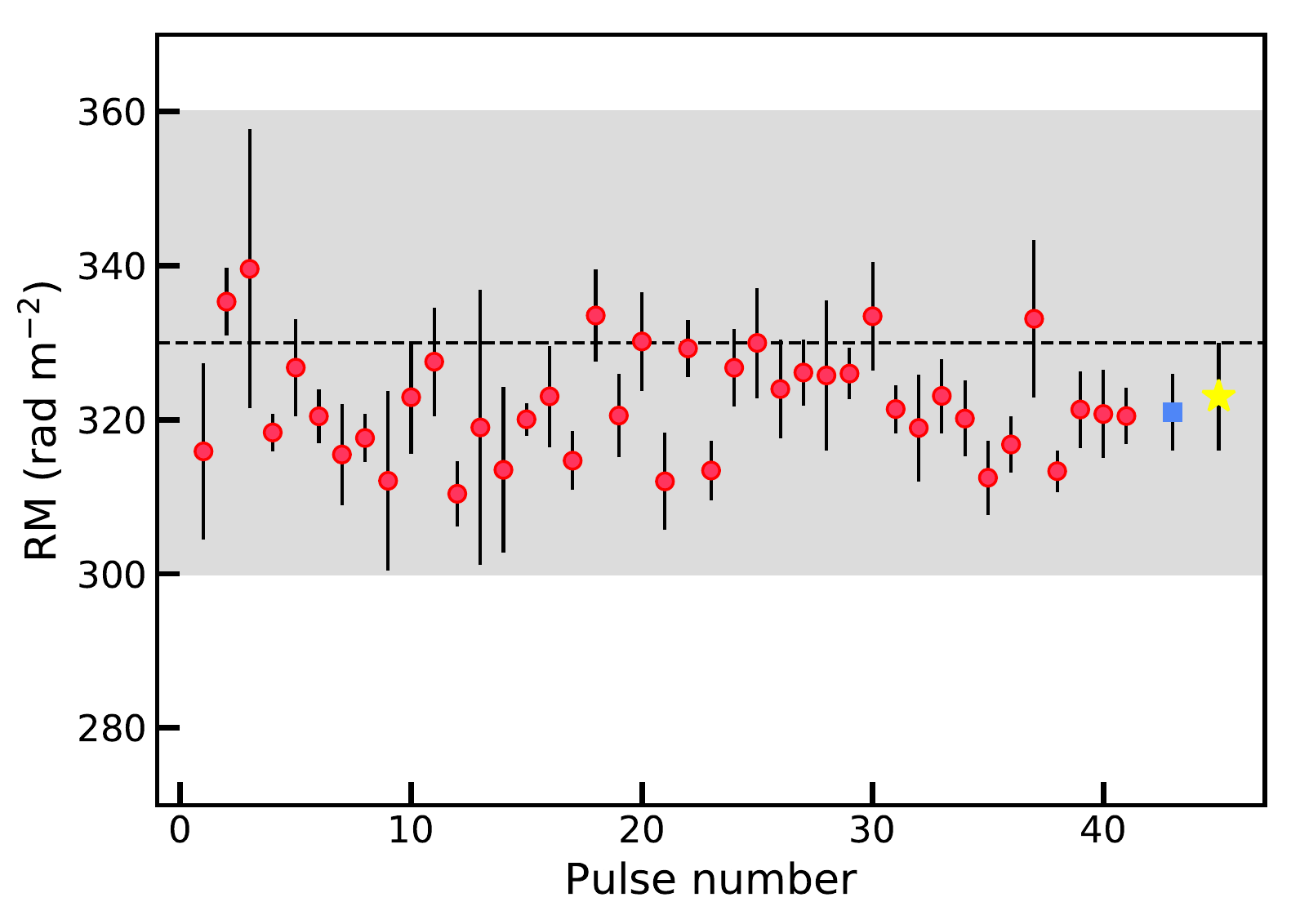}
\caption{RM as a function of pulse number for RRAT J1819$-$1458. The circles represent the RM of each single pulse whilst the star and square represent the RM of the weighted average of the single pulse distribution and the RM of the total integrated profile. The dashed line indicates the integrated pulse RM value measured by \citet{Karastergiou} and the shaded region its quoted uncertainty. }
\label{fig:J1819} 
\end{figure}  


\section{Results and analysis}
\label{sec:RandD}

Table \ref{table:params1} gives an overview of the results and our analyses. For each source, a distribution of the RMs of the single pulses was produced and an average RM value was determined, weighted by the inverse of the variance of each RM estimate. The single pulses are seen to exhibit variable degrees of linear and circular polarization, as seen in the normal pulsar population. The observed average degree of linear polarization for each RRAT is shown in Table \ref{table:params1} with individual single pulses being up to 100 per cent linearly polarized. The single pulses that yielded an RM value were grouped together (based on the day of observation)
and added to estimate the RM of the total integrated pulse.
A comparison of the RM of the integrated pulse with the weighted average of the single pulse RM distribution showed that the two are consistent to within a 2-$\sigma$ uncertainty. The degree of linear polarization in the integrated pulses of those exhibiting orthogonal modes (see Table \ref{table:params1}) is lower than the degree of linear polarization in the individual pulses, as the addition of orthogonally polarized radiation results in depolarization of the signal. 
Additionally the pulses from 5 RRATs, namely J0735$-$62, J0847$-$4316, J1047$-$58, J1854$+$0306 and J1909$+$06 exhibited insufficient linear polarization to obtain an RM measurement. Figures \ref{fig:profiles1} and \ref{fig:profiles2} show examples of polarized single pulses after RM correction. The pulses were chosen to show the diverse polarization properties exhibited by individual RRAT pulses (e.g. orthogonally polarized modes (see Section \ref{sec:conc}) and 100 per cent linear polarization). Since no flux calibrator observations were performed with the BPSR backend prior to each RRAT observation, the flux densities of the RRATs have arbitrary units.

To validate our method, we compare our results for RRAT J1819$-$1458, with the results of \cite{Karastergiou}. In Figure \ref{fig:J1819}, we compare the RM estimates of our single pulses and the RM of our integrated pulse with the integrated total value of $330 \pm 30$ rad m$^{-2}$ published in \cite{Karastergiou}. 
All the single pulse RMs from this work are found to be in good agreement with the published value.
We obtain a weighted average RM of $323 \pm 7\, \mathrm{rad \, m^2}$ from the single pulse distribution and a integrated pulse RM of $321 \pm 5 \, \mathrm{rad \, m^2}$.

\begin{figure}
\centering
\includegraphics[width=3.4 in]{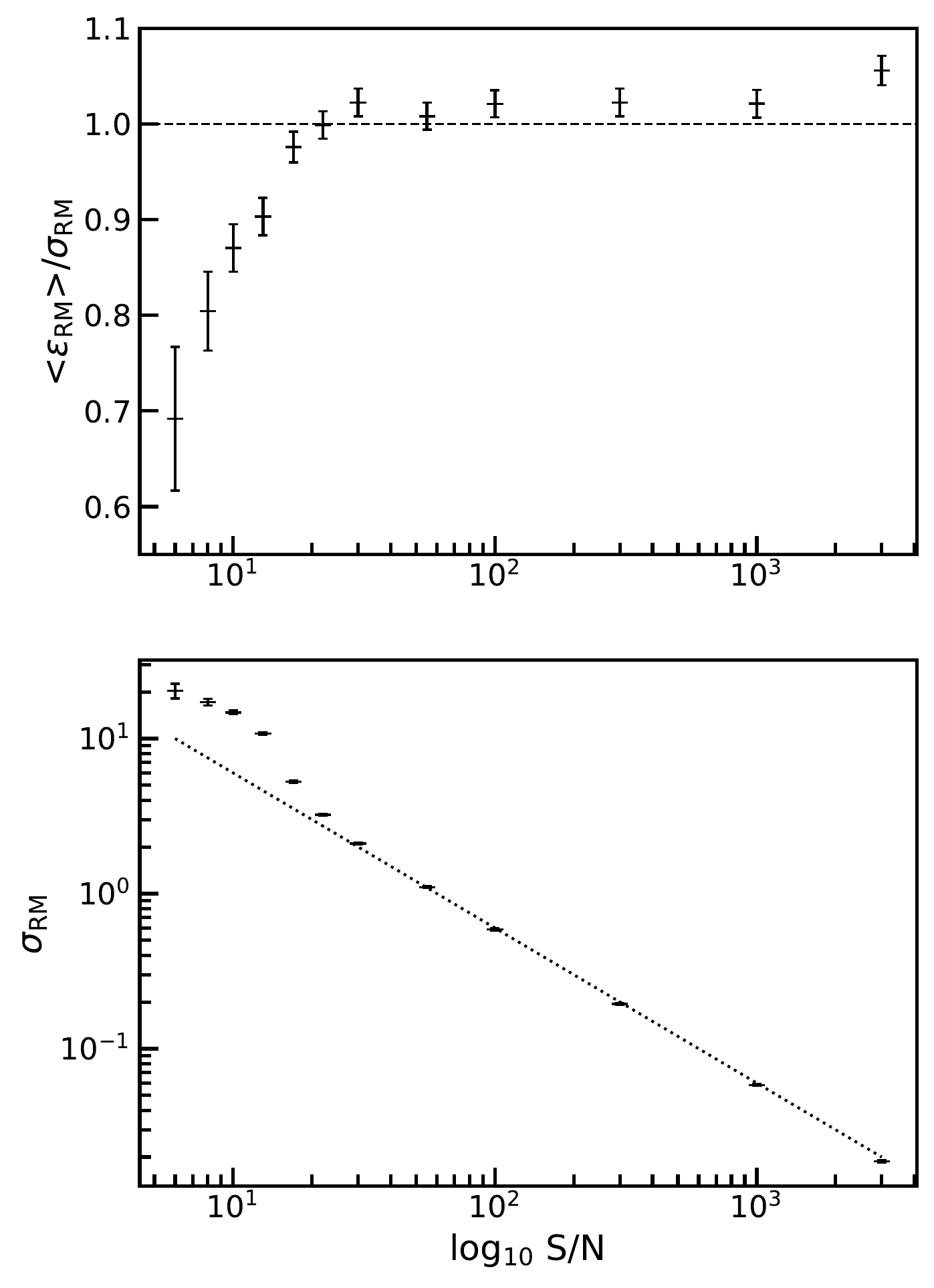}
\caption{Top panel: Ratio between the mean estimated uncertainty $\epsilon_\mathrm{RM}$ and the standard deviation of RM estimates $\sigma_\mathrm{RM}$ as a function of signal-to-noise ratio (S/N).  The expected ratio of unity is indicated by the dashed line. Bottom panel: S/N as a function of the standard deviation of the RM estimates $\sigma_\mathrm{RM}$. The standard deviation is expected to scale inversely with S/N as indicated by the line with slope $= -1$. The RM estimates output by \textsc{rmfit} do not follow this trend at low S/Ns.}
\label{fig:rmfit_error_ratio}
\end{figure}

To further verify the RM refinement algorithm described in the previous section and explore the minimum S/N required for a reliable RM estimate, we ran a Monte-Carlo simulation in which single pulses are generated with known RMs and S/Ns. For each trial value of S/N$_{i} \in \{6, 8, 10, 13, 17, 22, 30, 55, 100, 300, 1000, 3000\}$ and RM$_{i} \in \{\pm3000, \pm1000, \pm300, \pm100, \pm30, \pm10, \pm3, \pm1, 0\}$, 2500 artificial pulses are generated using {\sc psrsim} \citep{WvS_psrsim}. Each simulated pulse has a shape defined by the von Mises distribution with a concentration parameter $\kappa \sim 40$ and is sampled using 256 phase bins and 1024 frequency channels spanning a simulated bandwidth of 400~MHz centred at 1369~MHz \citep[i.e. matching the instrumental configuration of the HTRU survey described in][]{Keith}. 
In each phase bin, the degree of linear polarization is set to
90 per cent 
and normally distributed noise is added such that the S/N of the linearly polarized pulse profile after integrating over all frequency channels is equal to the target S/N$_i$.
Analysis of the results of this simulation verifies that the RM estimates are unbiased (i.e.\ $\langle$RM$\rangle$ - RM$_i$ is consistent with zero) for all trial RM and S/N values. For S/N $\geq 22$, the mean reported RM uncertainties $\langle\epsilon_\mathrm{RM}\rangle$ slightly over-estimate the standard deviations of RM estimation errors $\sigma_\mathrm{RM}$ (see Figure \ref{fig:rmfit_error_ratio}) and $\sigma_\mathrm{RM}$ is inversely proportional to S/N, as expected.  

For trials with S/N $< 17$, $\langle \epsilon_\mathrm{RM}\rangle$ underestimates $\sigma_\mathrm{RM}$ and the algorithm is successful in only 2.5 per cent of trials for S/N=6 and 37.7 per cent of trials for S/N=10. At low S/N, the uncertainty is underestimated because Equation~\ref{var_Delta_Psi} is correct only to first order.  The algorithm fails when linear polarization is not detected above 3$\sigma$ in any \emph{individual} phase bin, and the trial S/N$_i$ is that of the \emph{integrated} flux in all on-pulse phase bins; therefore, it is expected that some fraction of trials will fail at low S/N.
The limit that S/N must be $> 17$ is somewhat conservative because {\sc psrsim} simulates the linearly polarized flux using a Rotating Vector Model \citep[RVM;][]{Rad} that has a small impact angle of 6$^{\circ}$, such that there is a steep gradient in PA near the pulse peak (magnetic meridian). In principle, if the PA curve was flatter, it would be possible to further integrate and achieve greater S/N in each phase bin; this would enable {\sc rmfit} to succeed more often at lower S/N because the linearly polarized flux would exceed the 3$\sigma$ cut-off in a greater number of phase bins.

\begin{figure}
\centering
\includegraphics[width=3.2 in]{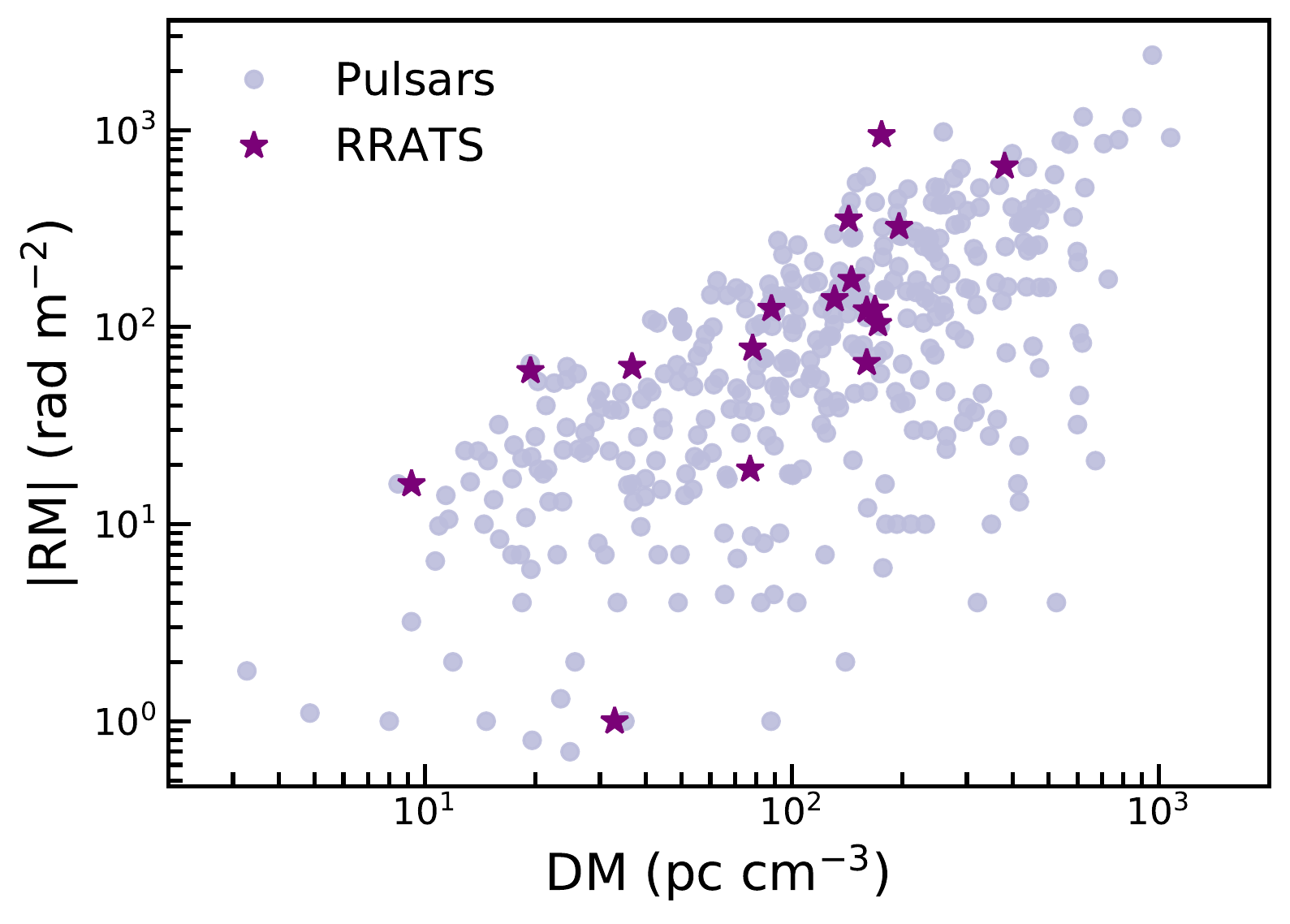}
\caption{The absolute RM values of the pulsars and RRATs in our Galaxy as a function of DM. Overall, the RMs of the RRATs are found to be consistent with those of the pulsar population.}
\label{fig:rms}
\end{figure} 

\begin{figure*}
\centering
\includegraphics[width=6.5 in]{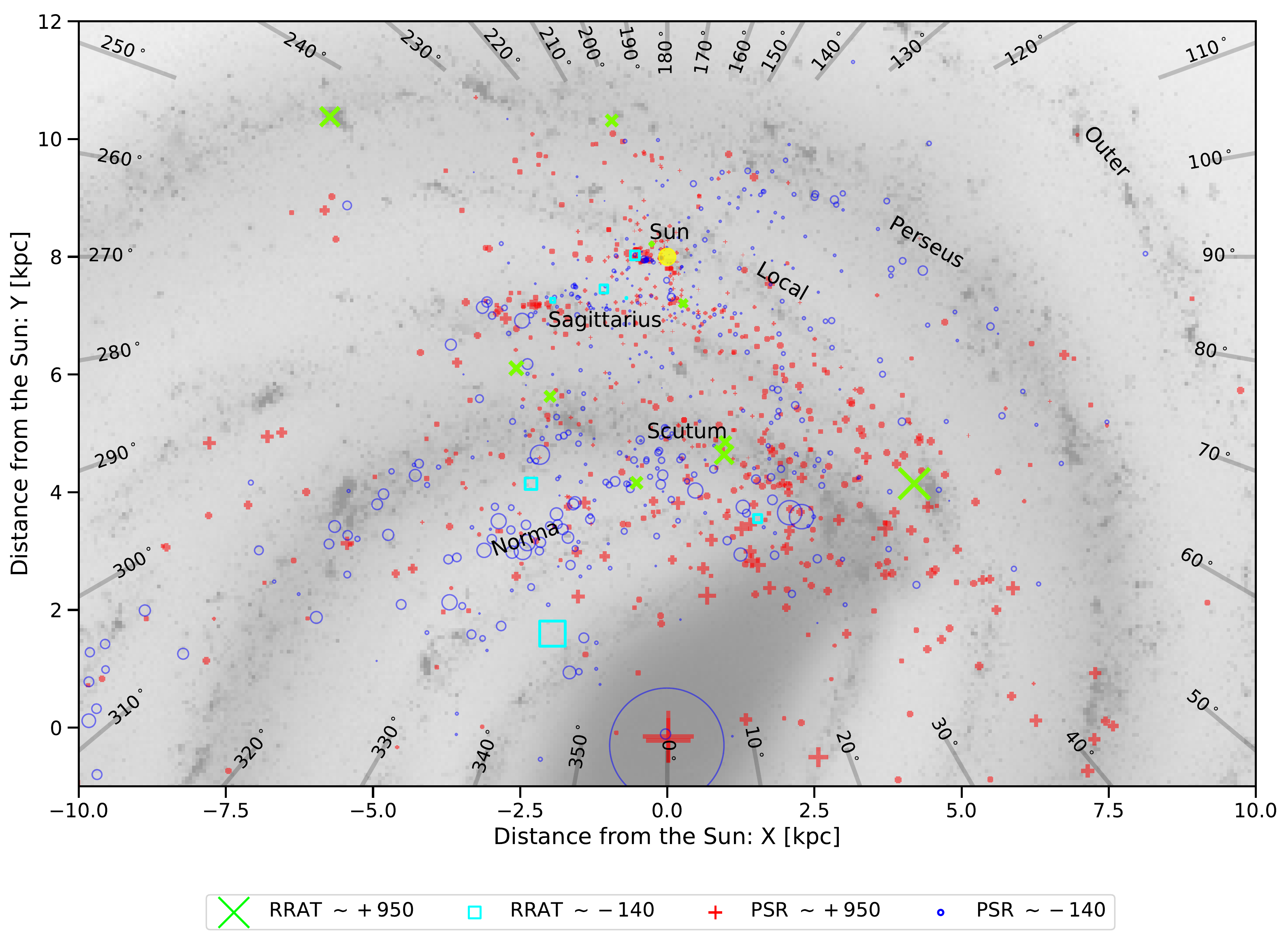}
\caption{Distribution of pulsars and RRATs with known RMs in our Galaxy. The crosses ($\times$) and pluses ($+$), and the squares and open circles represent the positive and negative RMs of the RRATs and pulsars respectively. The symbol size is proportional to the absolute RM. The linear sizes of the symbols are 10 times smaller in the case of pulsars and 2 times smaller in the case of the RRATs. Distance estimates to the pulsars and RRATs are based on the YMW16 model \citep{Yao}. The background is the image made by J.~Hunt of the Milky Way's disk, bulge/bar and spiral arms, as mapped out by the  GLIMPSE survey \citep{glimpse}.}
\label{fig:GPRMs}
\end{figure*}

\section{Discussion and Conclusions}
\label{sec:conc}

Figure \ref{fig:rms} displays the RMs of the pulsars and RRATs in our Galaxy as a function of their DMs. The values of the pulsar RMs 
were obtained from the ATNF pulsar catalogue \textsc{psrcat}\footnote{\url{http://www.atnf.csiro.au/people/pulsar/psrcat/}} \citep{DManchester} and the RMs of the 
RRATs are the weighted averages of the single pulse RM distributions from this work. We performed a two sample Kolmogorov-Smirnoff (KS) test on the RM distributions of the pulsars and RRATs with a $p$-value of $< 0.05$ being our criterion for deciding if the two distributions differ. We estimate a $p$-value of 0.4 which supports the hypothesis that the two distributions are drawn from the same population. This advocates for RRATs being related to the class of long-period pulsars, which were preferentially found in single pulse searches (until the Fast Folding Algorithm revolution) \citep{Staelin, Lovelace, Cameron, Tan}.
Of the $\sim 3200$ pulsars known, only $\sim 1120$ pulsars have measured RMs. We encourage recording full Stokes data for all single pulse detections (irrespective of whether it is deemed to be an FRB or not) above a S/N of 17 (assuming 90 per cent linear polarization), now that we have established that single pulse RMs can be reliably measured.
The line-of-sight averaged magnetic fields to the RRATs were determined by using the estimated RMs of the RRATs in Equation \ref{eq:Bll}.
The RM values and magnetic field strengths as a function of distance from the Galactic centre were found to be consistent with those measured using pulsars \citep[e.g.][]{Han2018} as shown in Figure \ref{fig:GPRMs}.

Polarization is a key diagnostic of radio emission and relates to emission geometry. Often, the position angle of the linearly polarized flux varies in a regular and smooth manner throughout the pulse -- a characteristic S-shaped curve explained by the Rotating Vector Model (RVM) \citep{Rad} -- and is independent of the observing frequency. However not all pulsars exhibit this behaviour. \cite{ManchesterOPM} show that several single pulses from pulsars show discontinuities in an otherwise unvarying PA swing resulting from the presence of two orthogonally polarized modes (OPMs). The classic signature of this effect is an abrupt jump of $\sim 90^{\circ}$ in the PA at a given pulse phase. Some RRATs are observed to exhibit OPMs in their single pulses as shown in Table \ref{table:params1}. We attempt to fit for the RVM model using a stable integrated pulse profile as opposed to an individual single pulse. However, the insufficient number of PA points in our integrated RRAT pulse profiles does not enable a robust fit for the RVM model.

In this paper we present the polarization analyses of 22 known RRATs with flux densities $S_{1400} >$100 mJy. Only 17 have sufficient linearly polarized flux (>3$\sigma$) in their single pulses to enable RM measurements. As a consistency check, we compared the published RM of J1819$-$1458 \citep{Karastergiou} with the weighted mean of our single pulse RM distribution and the RM of the integrated pulse profile, and both values were found to be consistent within the uncertainties. The RMs of the RRATs reported in this paper are found to be consistent with the RMs of the overall pulsar population. We verify the \textsc{rmfit} algorithm by simulating artificial single pulses over a wide range of RM and S/N. All trial runs succeeded in reporting accurate RM estimates for S/N $\geq 17$ in the linearly polarized flux, while only $37.7$ per cent and $2.5$ per cent of the trial runs succeeded for S/N = 10 and S/N = 6 respectively. As expected, the standard deviation of the RM estimates is observed to be inversely proportional to the S/N, however this trend fails at low S/Ns as seen in Figure \ref{fig:rmfit_error_ratio}.
We note that for pulses with S/N $\geq 22$ in linear flux, the average RM uncertainty reported by \textsc{rmfit} slightly over-estimates the standard deviation of the set of RM estimates, while the inverse is true for pulses with S/N $<17$.  
Future polarization measurements of more RRATs at different frequencies will help better understand if they are like normal pulsars as they should appear more polarized at low frequencies, and are probably completely unpolarized at higher frequencies. 
The method used for the RM estimations in this paper is similar to the one used to measure the RMs of individual FRB pulses \citep{Petroff_FRB, nat_keane, Caleb_poln} and validates their reliability.

\section*{Acknowledgements}
MC and BWS acknowledge funding from the European Research Council (ERC) under the European Union's Horizon 2020 research and innovation programme (grant agreement No 694745). CF acknowledges financial support by the Beckwith Trust. EP acknowledges support from an NWO Veni fellowship. Parts of this research were conducted by the Australian Research Council Centre
for All-Sky Astrophysics (CAASTRO), through project number
CE110001020 and the ARC Laureate Fellowship
project FL150100148. This work was performed on the ozstar national facility at Swinburne University of Technology. Ozstar is funded by Swinburne and the Australian Government's Education Investment Fund.

\bibliographystyle{mnras}
\bibliography{thesis}


\appendix

\section{RM refinement algorithm}
\label{sec:A1}

This appendix describes the RM refinement algorithm implemented by \textsc{rmfit} and first used in \cite{Han};
it is based on a similar algorithm introduced by \cite{hmq99}. During each step of iterative refinement, the algorithm computes the weighted differential position angle $\Delta\Psi$ and its uncertainty by cross-correlating two linear polarization pulse profiles, one integrated from the upper half of the band and the other integrated from the lower half of the band. Let
\begin{equation}
P_k = Q_k + i U_k = L_k \exp i2\Psi_k
\end{equation}
represent the linear polarization of the $k$th phase bin in the mean pulse profile associated with
radio wavelength $\lambda$ and
\begin{equation}
P^\prime_k = Q^\prime_k + i U^\prime_k = L^\prime_k \exp i2\Psi^\prime_k
\end{equation}
represent the linear polarization of the $k$th phase bin in the profile associated with
radio wavelength $\lambda^\prime$.
Consider the total cross-correlation,
$Z=C+iS$, where $C$ and $S$ are the sums of the real and imaginary components of
\begin{equation}
z_k = c_k + i s_k 
    = P^*_k P^\prime_k = L_k L^\prime_k \exp i2(\Psi^\prime_k - \Psi_k),
\end{equation}
\begin{equation}
c_k = Q_kQ^\prime_k + U_kU^\prime_k,
\hspace{5mm}\mathrm{and}\hspace{5mm}
s_k = Q_kU^\prime_k - U_kQ^\prime_k.
\end{equation}
\noindent
The weighted mean value of $\Psi^\prime_k - \Psi_k$,
\begin{equation}
\Delta\Psi = {1\over2}\tan^{-1}\left({S\over C}\right),
\end{equation}
and to first order the variance of this estimator,
\begin{equation}
{\mathrm{var}}(\Delta\Psi) = {1\over4}
{ C^2\sigma_{S}^2 + S^2\sigma_{C}^2
  + 2 SC \sigma_{S C} \over 
  \left(C^2 + S^2\right)^2 },
  \label{var_Delta_Psi}
\end{equation}
where the variances of the sums,
\begin{equation}
\sigma_{C}^2 = \sum_{k=1}^N
 Q^{\prime2}_k\sigma_{Q_k}^2 + U^{\prime2}_k\sigma_{U_k}^2 +
 Q^2_k\sigma_{Q^\prime_k}^2 + U^2_k\sigma_{U^\prime_k}^2,
\label{eqn:var_ck}
\end{equation}
\begin{equation}
\sigma_{S}^2 = \sum_{k=1}^N
 U^{\prime2}_k\sigma_{Q_k}^2 + Q^{\prime2}_k\sigma_{U_k}^2 +
 U^2_k\sigma_{Q^\prime_k}^2 + Q^2_k\sigma_{U^\prime_k}^2,
\label{eqn:var_sk}
\end{equation}
and the covariance between them
\begin{equation}
\sigma_{C S} = \sum_{k=1}^N
 Q^\prime_k U^\prime_k (\sigma_{U_k}^2 - \sigma_{Q_k}^2) +
 Q_k U_k (\sigma_{Q^\prime_k}^2 -\sigma_{U^\prime_k}^2).
\label{eqn:covar_ck_sk}
\end{equation}
Finally, the change in RM required to negate $\Delta\Psi$ is given by
\begin{equation}
    \Delta\mathrm{RM}={\Delta\Psi\over{{\lambda^\prime}^2 - \lambda^2}}.
\end{equation}

\label{lastpage}
\end{document}